\documentclass[sigconf]{acmart}

\usepackage{multirow}
\usepackage{stfloats}
\usepackage{subfigure}
\usepackage{makecell}
\AtBeginDocument{%
  \providecommand\BibTeX{{%
    \normalfont B\kern-0.5em{\scshape i\kern-0.25em b}\kern-0.8em\TeX}}}

\setcopyright{acmcopyright}
\copyrightyear{2021}
\acmYear{2021}
\acmDOI{10.1145/1122445.1122456}

\acmConference[SUI '21]{SUI '21: ACM Symposium on Spatial User Interaction}{November 09--10}{2021}
\acmBooktitle{SUI '21: ACM Symposium on Spatial User Interaction,
  November 09--10, 2021}
\acmPrice{15.00}
\acmISBN{978-1-4503-XXXX-X/18/06}

\begin{document}

\title{Effect of Visual Cues on Pointing Tasks in Co-located Augmented Reality Collaboration}

\author{Lei Chen}
\email{lei.chen02@student.xjtlu.edu.cn}
\affiliation{%
  \institution{Xi'an Jiaotong-Liverpool University}
  \city{Suzhou}
  \country{China}
}

\author{Yilin Liu}
\email{yilin.liu1999@gmail.com}
\affiliation{%
  \institution{Xi'an Jiaotong-Liverpool University}
  \city{Suzhou}
  \country{China}}

\author{Yue Li}
\email{yue.li@xjtlu.edu.cn}
\affiliation{%
  \institution{Xi'an Jiaotong-Liverpool University}
  \city{Suzhou}
  \country{China}
}

\author{Lingyun Yu}
\email{lingyun.yu@xjtlu.edu.cn}
\affiliation{%
  \institution{Xi'an Jiaotong-Liverpool University}
  \city{Suzhou}
  \country{China}
  }
  
\author{BoYu Gao}
\email{bygao@jnu.edu.cn}
\affiliation{%
  \institution{Jinan University}
  \city{Guangzhou}
  \country{China}}

\author{Maurizio Caon}
\email{maurizio.caon@hes-so.ch}
\affiliation{%
  \institution{University of Applied Sciences and Arts Western Switzerland (HES-SO)}
  \city{Fribourg}
  \country{Switzerland}}

\author{Yong Yue}
\email{yong.yue@xjtlu.edu.cn}
\affiliation{%
  \institution{Xi'an Jiaotong-Liverpool University}
  \city{Suzhou}
  \country{China}}

\author{Hai-Ning Liang}
\email{haining.liang@xjtlu.edu.cn}
\authornote{Corresponding author (haining.liang@xjtlu.edu.cn)}
\orcid{0000-0003-3600-8955}
\affiliation{%
  \institution{Xi'an Jiaotong-Liverpool University}
  \city{Suzhou}
  \country{China}}

\renewcommand{\shortauthors}{Chen, et al.}

\begin{abstract}
 Visual cues are essential in computer-mediated communication. It is especially important when communication happens in a collaboration scenario that requires focusing  several users' attention on a specific object among other similar ones. This paper explores the effect of visual cues on pointing tasks in co-located Augmented Reality (AR) collaboration. A user study (N = 32, 16 pairs) was conducted to compare two types of visual cues: Pointing Line (PL) and Moving Track (MT). Both are head-based visual techniques. Through a series of collaborative pointing tasks on objects with different states (static and dynamic) and density levels (low, medium and high), the results showed that PL was better on task performance and usability, but MT was rated higher on social presence and user preference. Based on our results, some design implications are provided for pointing tasks in co-located AR collaboration.
\end{abstract}

\begin{teaserfigure}
  \includegraphics[width=\textwidth]{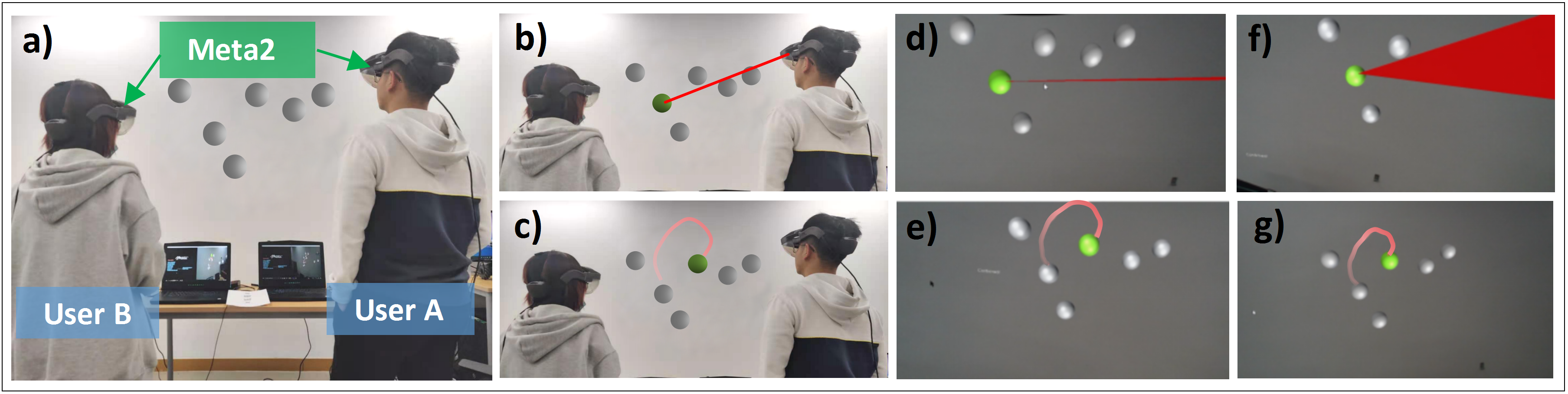}
 \caption{\textbf{(a)} A picture of the experimental setup; Two users are locating a target using \textbf{(b)} Pointing Line (PL) cues and \textbf{(c)} Moving Track (MT) cues; The view of User A who is locating a target using \textbf{(f)} PL and \textbf{(g)} MT; The view of User B who is now looking the target that User A has selected using \textbf{(d)} PL and \textbf{(e)} MT.}
  \label{Fig1}
\end{teaserfigure}

\maketitle

\section{Introduction}
Augmented Reality (AR) has been used for collaborative experiences. It allows users to interact with shared virtual content while having a view of the real world \cite{lukosch2015,billinghurst2002,yamazoe2014}. For these collaborative experiences to be efficient and positive, there needs to be fluid communication between the collaborators. In addition, knowing what the collaborator is doing or which objects are being looked at can improve the sense of awareness and social presence \cite{kim2018effect,kim2019evaluating,gerbaud2008scenario,kraut1996collaboration}. One advantage of using AR head-mounted displays (HMDs) is their ability to capture users' head and gaze movements and use the data to provide visual augmentation cues \cite{erickson2020sharing,huang2018handsintouch, XuPointing2019}. Such cues can enhance situational awareness and social presence \cite{bai2020user} and improve user performance and usability in collaborative AR  \cite{piumsomboon2019effects}. 

Previous research on AR primarily focused on remote collaboration (e.g., \cite{piumsomboon2017covar, higuch2016can, huang2018augmented, huang2018handsintouch}). It is not clear how visual cues could be helpful in enhancing user collaboration when they are in the same physical environment. For pointing tasks specifically, many applications involve such pointing tasks during co-located collaboration (e.g., games, education, training). It is therefore important to understand the appropriate techniques for such pointing tasks with different object states and density levels. Users could be interacting with objects that are not easily touchable (e.g., moving objects) and clustered in dense regions with other objects with similar properties. Attempting to describe or pinpoint an object of interest using verbal and hand gestures may not be practical and efficient. In this research, we explore the use of visual cues to enhance the identification of target objects in collaborative AR where these objects can be static or dynamic and be in various levels of density.

Pointers \cite{erickson2020sharing,kim2013comparing, yuEndPoint2019} and annotations \cite{huang2019sharing,laviola1998collaborative} are two main visual cues that have been explored. The majority of prior research focused on hand-based techniques, e.g., to carry a pointer and to draw an annotation in a virtual environment. However, these hand-based techniques may not be efficient or ideal for AR systems \cite{xue2021itext, LuHandfree2020}. For one, it can pose severe occlusion because the field-of-view of AR HMDs is usually narrow. Also, AR systems, with the exception of the Magic Leap, do not come with a pointing device or controllers. 

Therefore, in this research, we explore two hands-free, head-based techniques to simulate visual cues provided by pointers and annotations. Pointing Line (PL) is used to simulate pointers. It indicates a user's line of sight and focus of attention. Moving Track (MT) simulates the use of annotations. It records a continuous moving track to help identify a target. Based on the two techniques, we investigate the effect of visual cues on task performance, usability, and social presence when interacting with virtual objects in a co-located AR. To do this, we ran a user study with two visual cues (PL and MT) to allow paired users to share information about the object of interest and identify it. The experiment involved both static and dynamic objects in three levels of density (low, medium, and high). Figure~\ref{Fig1} (above) shows an overview of the experiment setup. 

In short, the paper makes the following two contributions:
\begin{itemize}
  \item We report the results of a user study comparing the two types of visual cues based on hands-free, head-based techniques in co-located AR scenarios for pointing and selection tasks; 
  \item Based on the results, we propose implications on the design and use of these visual cues for co-located AR collaboration. 
\end{itemize}

\section{Related Work}
\subsection{Collaborative AR Systems and Social Presence}
Collaboration is the “mutual engagement of participants in a coordinated effort to solve a problem together” \cite{roschelle1995construction}. 
Collaborative AR systems allow users to interact with shared AR content as naturally as with physical objects, and to complete a task or achieve a common goal \cite{billinghurst1999collaborative,li2017multi,chen2020collaborative}. 
Researchers have shown that AR systems can effectively support a group of users to perform collaborative activities \cite{tait2015effect,kaufmann2003collaborative,szalavari1998collaborative,billinghurst2002experiments}.

During collaboration, it is key to be aware of where and which object(s) the collaborator is interested in or interacting with \cite{lacoche2017collaborators,antunes2001abstraction,dourish1992awareness}. Providing visual, nonverbal cues is beneficial for improving the awareness of users' actions and the sense of being together, namely social presence \cite{buxton2009mediaspace,gergle2013using}. In the early days, telepointers and cursors \cite{greenberg1996semantic} were explored to support user awareness of others’ actions on a shared workspace in traditional platforms (e.g., desktop displays). Recently, multimodal cues are used in collaborative environments, typically using auditory and visual elements. For example, virtual avatars  have been explored to represent each collaborator and to provide an increased awareness of others in the shared environment  \cite{gutwin1996workspace,piumsomboon2018mini,jo2016effects}. In AR environments, researchers explored the use of virtual arrows to represent collaborators' head directions \cite{chastine2007understanding} and miniature virtual avatars to show collaborators' gaze directions and body gestures \cite{piumsomboon2018mini}. Although using avatars can contribute to users' perceived social presence, it adds extra visual elements to the limited display and field-of-view of current AR HMDs. Thus, this solution may not be ideal when there are multiple objects in the environment. Using pointer and annotation cues is an uncluttered alternative to support social presence and collaboration in AR. Specifically, sharing gaze pointing cues has been explored in remote collaboration with AR devices \cite{ishii1993integration, lee2017improving, higuch2016can}. For example, Ishii et al. \cite{ishii1993integration} and Higuchi et al. \cite{higuch2016can} reported that when users shared their workspace, gaze pointing cues provided better understanding where their partner was looking at. Lee et al. \cite{lee2017improving} and Higuchi et al. \cite{higuch2016can} found that sharing gaze pointing cues significantly improved users' awareness of each other's focus and joint attention. 

In addition to gaze pointing cues, augmenting hand pointing cues via gestures was shown to be able to facilitate users' perception of others' actions. For example, Piumsomboon et al. \cite{piumsomboon2018mini} reported that redirected gaze and gestures in their Mini-Me system improved users' awareness of the partner in a collaborative AR interface. Yang et al. \cite{yang2020effects} stated that visual head frustum and hand gestures intuitively demonstrated the remote user’s movements and target positions. Kim et al. \cite{kim2019evaluating} added sketch cues in addition to gestures and demonstrated an improved task efficiency with the enhanced visual cues. 

Although previous research has proven the positive effects of visual cues on supporting social presence in collaborative activities, most of these studies focused on remote AR systems. The effect of visual cues in co-located AR collaboration is still largely underexplored. It is not clear how different visual cues may enable and enhance users' perceived social presence when they are co-located in the same physical space. Therefore, we put our focus on the study of visual cues in co-located AR environments.
\subsection{Visual Cues in Collaboration}
Two noticeable visual cues were presented in previous work on collaboration: pointer and annotation \cite{kim2013comparing, huang2019sharing, teo2018hand, erickson2020sharing}. Pointer cues provide a pointing line, indicating a user's line of sight and focus of attention. Annotation cues record a moving track to identify a target, such as a track of hand or head positions. Here, we discuss these two types of visual cues in detail.

\paragraph{\textit{\textbf{Pointing Line Cues}}}Previous work showed that pointing line cues can effectively support communication \cite{greenberg1996semantic,duval2014improving,oda2015virtual,sousa2019warping,sakata2003wacl,gupta2016you}. For example, Gupta \cite{gupta2016you} reported that presenting users' gaze directions using a pointer significantly improved the sense of co-presence between users in remote collaboration. Piumsomboon et al. \cite{piumsomboon2019effects} visualized three types of pointing line cues: the field-of-view frustum, eye-gaze ray, and head-gaze ray. They reported that these cues significantly improved user performance, usability, and subjective preferences. They also found that head-gaze ray was significantly less confusing to use than field-of-view. 

Most of the previous research explored the effect of pointing line cues with hand gestures or eye gaze \cite{bai2020user,sousa2019warping}. For hand gestures, if users' hands are occupied, it requires users to deliberately release the object in their hand before pointing to a target. Some research used eye gaze to provide pointing line cues \cite{vspakov2019eye,blattgerste2018advantages}. They stated that eye gaze can lead to better performance and better teamwork than head gaze, i.e., the direction that a user is facing towards but not necessarily looking at. However, people's eye gaze has a lot of micro movements. Having to keep their eyes stable when pointing at an object may lead to fatigue. Head-based techniques, on the other hand, could mitigate this issue. By providing a gaze ray from a user’s head to a target object, this can support observers' awareness of their collaborator’s attention and allow them to view the same object \cite{anthes2005toolbox}. However, studies on head-based pointing line cues are very limited, especially in co-located AR collaboration.

\paragraph{\textit{\textbf{Moving Track Cues}}}Moving track is one of the mostly studied visual communication cues \cite{tang1991videodraw,rekimoto1995world}. 
It has been found to be more effective than pointing line cues for communicating spatial information \cite{kim2013comparing,fussell2004gestures}. While the pointing line cues present a point around the target object, moving track cues provide users with a track, namely a series of points leading to the target, and can help guide users' focus to key locations \cite{xue2021itext}. Early research has studied the use of moving track cues on a shared video view \cite{tang1991videodraw,ishii1994iterative}. Researchers have explored extensively the use of moving track cues in remote collaborative physical tasks. For example,
Billinghurst et al. \cite{billinghurst2002real} presented a method to record stabilized moving tracks in an annotation-based AR system.
Teo et al. \cite{teo2018hand} reported that local users in remote collaboration felt the task was easier and more enjoyable with moving track cues, and the cues helped them understand the attention and focus of the remote user. A recent study \cite{kim2019evaluating} compared four combinations of visual cues provided by hand gesture, pointer and sketch. They reported that participants completed tasks faster and felt higher level of usability if sketch (moving track) cues were provided. 

Although there have been studies on moving track cues, such as annotations and drawings, this research mainly used hand-based techniques and focused on remote collaboration \cite{huang2018handsintouch,teo2018hand}. Besides, previous research on moving track cues also demonstrated some limitations. Much of this research explored these cues as an annotation tool. In this case, the track traces needed to be erased after completing each step of a task \cite{kim2013comparing}. Huang et al. \cite{huang2018handsintouch} also mentioned that moving track cues that blocked users' view could have a negative impact on user experience and task performance. Based on these findings, we have made the moving track cues to gradually fade away and disappear in our study. This solution neither requires users to actively erase any virtual element, nor prevents users from viewing the workspace with additional visual elements. 

In summary, our research is framed on previous work and compares two head-based visual techniques, pointing line cues and moving track cues, to explore their effect on pointing tasks in co-located AR collaboration.

\section{System Setup}
To conduct this research, we created a multi-user AR collaborative system (see Figure~\ref{Fig1}a). The system was developed with Unity (version 2019.2.0f1) and was run on the Windows platform. Each of the two users used a Meta2 connected to a laptop and with each other via a private network and Photon Unity Networking. Either laptop could be the host server. After they have been connected, one would assume the host role and the two laptops’ timers were synchronized to begin the experiment.

The equipment set up in this study consisted of (1) two Meta2 HMDs, an AR device with two optical see-through displays with a 2550 x 1440 pixel resolution, 90° field of view, and frame rate of 60 Hz. (see Figure~\ref{Fig1}a), and (2) two Windows 10 laptops each with an Intel Core i9-8950HK at 2.9 GHz, 32 GB RAM, and NVIDIA GeForce GTX 1080. This set up allows a pair of users to work in the same virtual environment using the two Meta2 HMDs and jointly do any tasks together. The system is able to collect data (e.g., completion time and accuracy).

\subsection{Visual Techniques}
To determine the techniques selected for this study, we first conducted a pilot study and explored both head-based and hand-based approaches, each with and without a visual cue (i.e., baseline). The results showed that head-based tended to be better in performance and user preference for pointing tasks in co-located AR (which was in line with other studies (e.g.,~\cite{hansen2004gaze,yang2020effects}). Given our pilot study results and those from the literature, we did not include a baseline technique to keep the study focused and within a reasonable time. Therefore, our study is based on two head-based visual techniques (i.e., Pointing Line and Moving Track) to compare their performance and usability under different scenarios.

The two head-gaze techniques receive input from the user's head movement and is captured by the tracking function in the Meta2. The cursor endpoint is based on a ray cast from the head’s center toward the AR interface.

\begin{figure*}[h]
  \centering
  \includegraphics[width=0.85\textwidth]{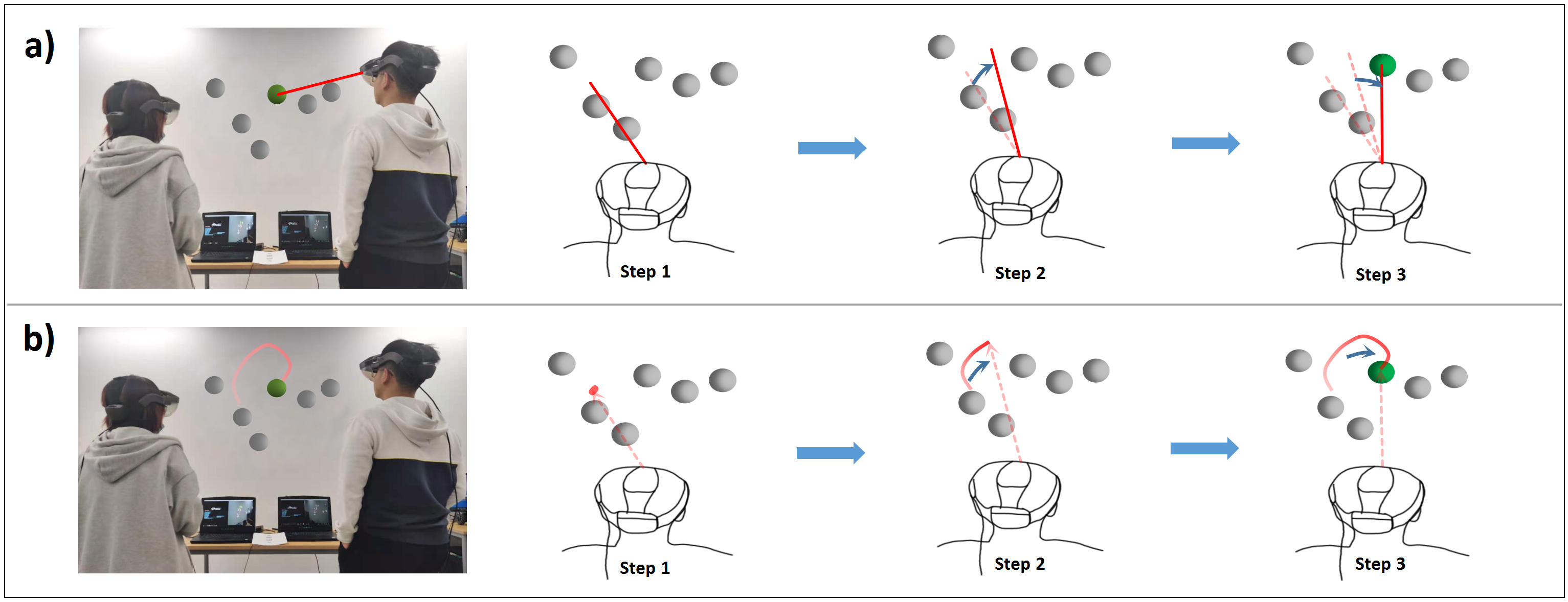}
  \caption{The two visual techniques \textbf{(a)} Pointing Line (PL) and \textbf{(b)} Moving Track (MT) with an example each.}
  \label{Fig2_}
\end{figure*}

\paragraph{\textbf{Pointing Line (PL)}}  PL is shown as a red ray representing the user’s viewing direction. It is emitted from the center of the user’s head toward the user interface in AR environment. When it intersects with an object, this object becomes highlighted and can be selected. As shown in Figure~\ref{Fig2_}a (upper row), when a user moves his/her head, the ray also moves in synchronicity following the head's directional movements. 

\paragraph{\textbf{Moving Track (MT)}}  MT uses a different approach to show directional movement. Instead of showing the ray, it shows a trail that follows the cursor. To reduce excessive visual clutter, the technique limits the length of the trail to 10 pixels. That is, the excess part of the tail fades out and gradually disappears. As shown in Figure~\ref{Fig2_}b (lower row), as the user moves their head, the trail is displayed, with end part becoming lighter and fading out gradually.

\section{USER STUDY}
\subsection{Participants}
16 pairs of participants (17 males) between the ages of 18 to 23 (M = 20.78) were recruited from a local university for this experiment. Only 3 pairs did not know each other before participating in the experiment. All participants had normal or corrected-to-normal vision and had no issues distinguishing the colors we used in the AR device. They reported an average of 3.69 for cooperative ability on a scale from 1 (‘so bad’) to 5 (‘good’). 13 of them (41\%) had limited prior experience with AR HMDs before. 

\subsection{Experiment Design}
\begin{figure}[h]
  \centering
  \includegraphics[width=0.45\textwidth]{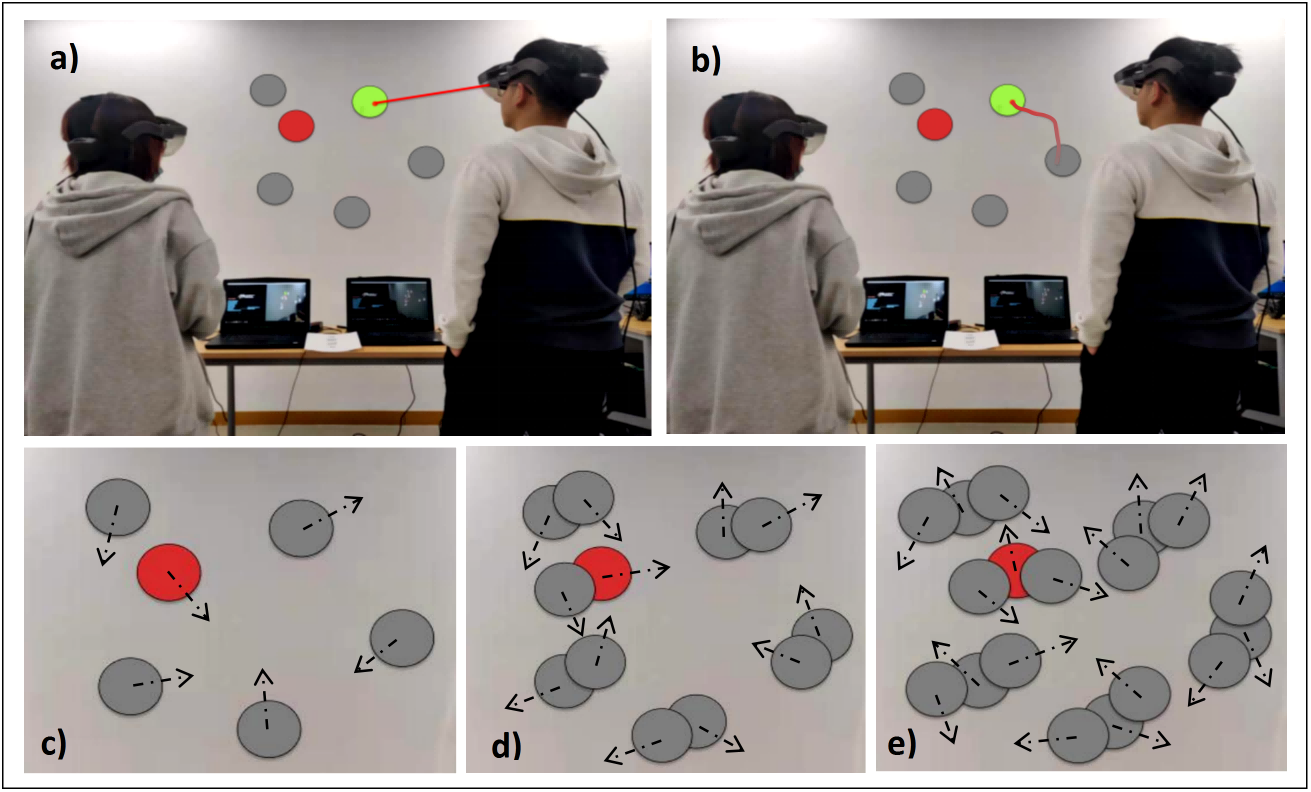}
  \caption{Two users completing the task with \textbf{(a)} Pointing Line (PL) and \textbf{(b)} Moving Track (MT) technique. There were \textbf{(c)} Low, \textbf{(d)} Medium and \textbf{(e)} High density for Static and Dynamic objects. Red indicates the target object; green indicates the object selected.}
  \label{Fig2}
\end{figure}

The experiment followed a 2 × 2 × 3 (2 Technique × 2 Object State × 3 Density) within-subjects design to study the effects of visual cues on pointing tasks in co-located AR collaboration. In this experiment, we have the following three independent variables: 
\begin{enumerate}
\item \textit{Technique}. We tested two visual techniques: \textit{Pointing Line} (PL) and \textit{Moving Track} (MT). As mentioned above, PL is based on the ray-casting approach to show a line to the object on the direction of the AR HMD (see Figure~\ref{Fig2}a). MT provides a continuous trace that follows the pointer (see Figure~\ref{Fig2}b).
\item \textit{Object State}. We explored two states: \textit{Static} and \textit{Dynamic}. Static objects were immobile while the dynamic ones would move around. As the arrows depicted in Figure~\ref{Fig2}c indicate, dynamic objects would be randomly assigned a path without any particular pattern. In dynamic scenarios, the objects performed uniform rectilinear motion (i.e., constant velocity) with speeds between 50-120 pixels/second and random directions. The placement and movement of the objects were predefined to be same for each pair. All objects (target and non-targets) would move and were programmed to avoid each other. A virtual boundary of 120 degrees was set for the objects. As this is wider than the 90-degree FOV of the Meta2, users would need to turn their heads during the tasks.
\item \textit{Density}. The size of objects used in our study was 70 pixels in diameter. There were three levels of object density: \textit{Low} (6 objects with no occlusion, Figure~\ref{Fig2}c), \textit{Medium} (12 objects with slight occlusion, Figure~\ref{Fig2}d) and \textit{High} (18 objects with severe occlusion, Figure~\ref{Fig2}e). 
\end{enumerate}

As a within-subjects experiment, each participant experienced all 12 conditions. Technique was counterbalanced according to a balanced Latin square design to minimize learning effects. Object State and Density were randomly and equally distributed. There were 6 trails for each condition and 72 trials in total for each pair of participants.

This experiment was classified as low risk research and was approved by the University Ethics Committee at Xi'an Jiaotong-Liverpool University (\#21-01-09).

\subsection{Hypotheses}
Based on our literature review and design of our experiment, we postulated the following four hypotheses:

\textit{\textbf{H1:}} PL and MT would both perform better with static targets than with dynamic objects.  

\textit{\textbf{H2:}} PL and MT would have better performance in lower density scenarios. High density situations could be crowded and have object occlusion, making it more difficult to identity and select a target. 

\textit{\textbf{H3:}} PL's more explicit cue showing viewing direction, it would lead to improved task efficiency and accuracy than MT.

\textit{\textbf{H4:}} In terms of subjective measures, in general, MT would be more preferred by users than PL, as MT would make it easier to track cursors movement and the user's intention (in our case User A). As such, it would increase collaborators’ awareness and co-presence.

\begin{figure*}[t]
  \centering
  \includegraphics[width=0.95\textwidth]{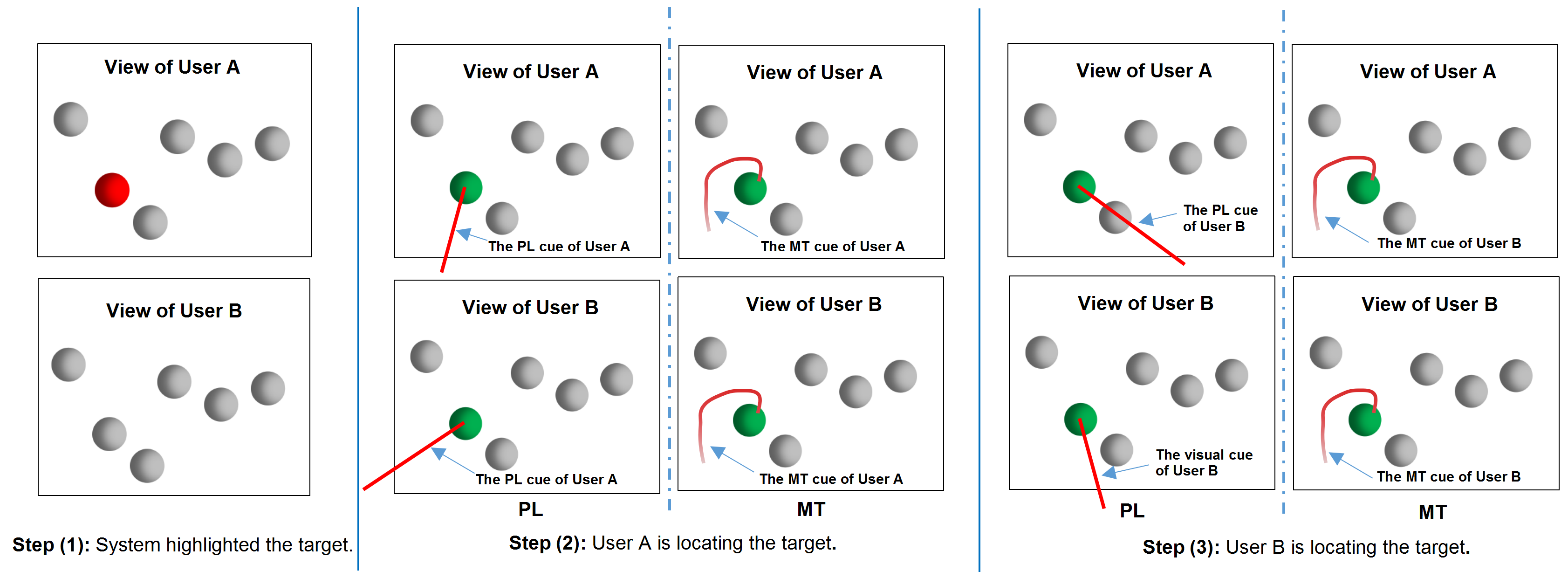}
  \caption{An example of all the steps involved in a typical trail.}
  \label{Figp}
\end{figure*}

\begin{table*}[b]
\centering
  \caption{Visibility states of the visual cues of User A and User B and the target to be selected according to selection stage}
  \label{tab3}
  \begin{tabular}{m{1cm}<{\centering}  p{1.25cm}<{\centering}   p{1.25cm}<{\centering}  p{3.4cm}<{\centering}   p{1.25cm}<{\centering}   p{1.25cm}<{\centering}  p{5.6cm}<{\centering} }
    \toprule
    \multirow{2}*{}& \multicolumn{3}{c}{\textbf{During the selection of User A}} & \multicolumn{3}{c}{\textbf{During the selection of User B}}\\
    \cline{2-7}
    &\multicolumn{1}{m{1.25cm}}{User A’s visual cue}&\multicolumn{1}{m{1.25cm}}{User B’s visual cue}&Target to be Selected&\multicolumn{1}{m{1.25cm}}{User A’s visual cue}&\multicolumn{1}{m{1.25cm}}{User B’s visual cue}&Target to be Selected\\
     \hline
     \textbf{User A}& {Yes}&{No}& \multicolumn{1}{m{3.2cm}}{Highlighted by System}& {No}&{Yes}&\multicolumn{1}{m{5.2cm}}{Visible (Already Selected by this User)}\\
      \hline
     \textbf{User B}& {Yes}&{No}& \multicolumn{1}{m{3.4cm}}{Visible but not highlighted}& {No}&{Yes}&\multicolumn{1}{m{5.6cm}}{Visible and Highlighted (Selected by User A)}\\
  \bottomrule
\end{tabular}
\end{table*}

\subsection{Task and Procedure}
\paragraph{\textit{\textbf{Task}}} For each trial, User A was required to choose the target object (see Step (1) in Figure~\ref{Figp}, highlighted in red) by moving the cursor to the object using head motions and confirming it via a mouse click. The selected object would turn to green, which can be seen by User B. Then User B would need to locate the same object by moving the cursor and doing the confirmation. As such, the task consisted of the following steps (see Figure~\ref{Figp}): (1) system randomly highlighted in red one object in the scene. Only User A could see the highlighted target. At this moment, the objects in User B's view are all grey. (2) User A would choose the object using the pre-defined technique. Once confirmed, the selected object would turn into green and both users could see this object. (3) User B was then required to locate the green object and confirm it. (4) One trial was finished and the next would begin when both users were ready to proceed. 

In our experiment, the two users in each pair would do the selection in turns. We designed the experiment so that they could not see both the visual cues of their partner and their own at the same time. During selection, each user could only see one visual cue (the cue of the user needing to do the selection). In other words, during the selection of User A, only the cue of User A would be visible; while during the selection of User B, only the cue of User B would be shown. Besides, at the beginning, only User A could see the target highlighted by system. Only after User A completed the selection would User B be able to see the selected target, now selected by User A (see the Table \ref{tab3} for a summary of this process). This approach was chose to minimize visual clutter and to make it more aligned with how collaboration takes in places in more realistic scenarios.

\paragraph{\textit{\textbf{Procedure}}} The experiment procedure was divided into 4 phases: (\textit{P1}) Informing participants of the experiment goal and the ethics regulations governing it, and then completing the consent form plus a short questionnaire to collect anonymized demographic data ($\sim$5 minutes); (\textit{P2}) Providing participants several practice trials to let them become familiar with the AR device and the task ($\sim$5 minutes); (\textit{P3}) Participants completing the trials and in between conditions each participant filling in the Social Presence, and Usability questionnaires ($\sim$25 minutes); and (\textit{P4}) Interviewing participants to collect further feedback and comments ($\sim$5 minutes). The whole experiment took about 40 minutes to complete for each pair.

\begin{figure*}[t]
  \centering
  \includegraphics[width=0.92\textwidth]{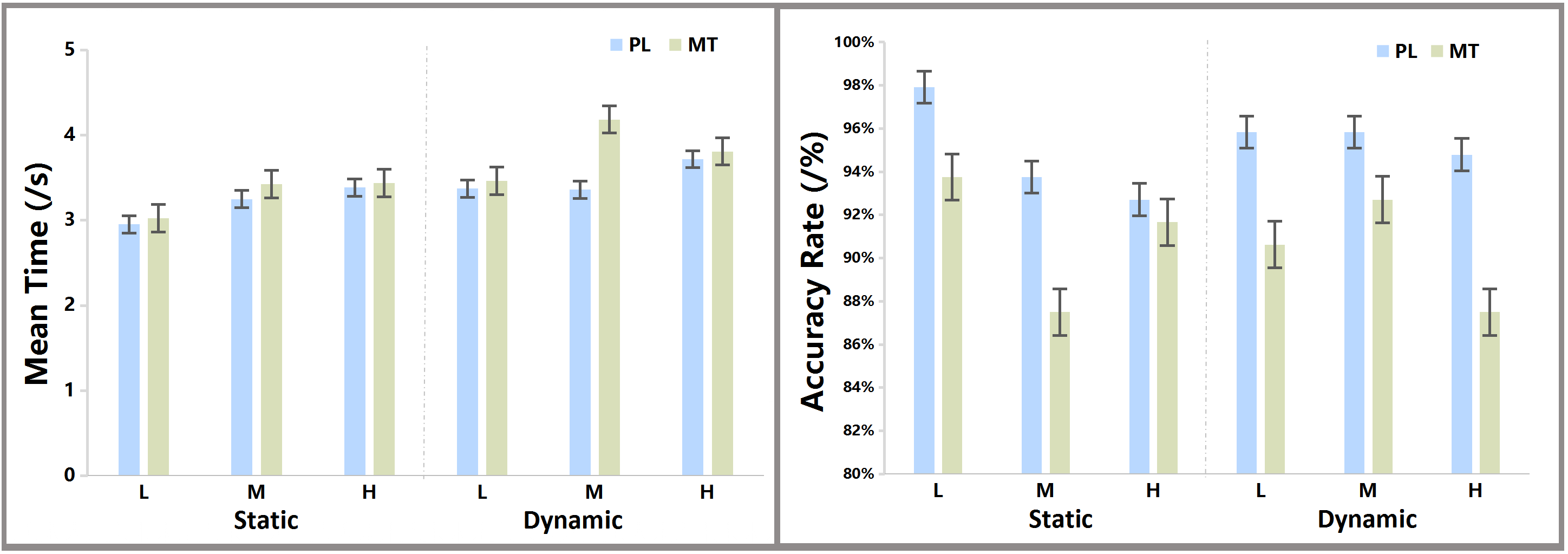}
  \caption{Mean completion time (Left) and accuracy rate (Right) results in all conditions: 2 Technique (PL and MT) × 2 Object State (Static and Dynamic) × 3 Density (L: Low, M: Medium and H: High).}
  \label{Fig3}
\end{figure*}
\begin{figure*}[b]
  \centering
  \includegraphics[width=0.95\textwidth]{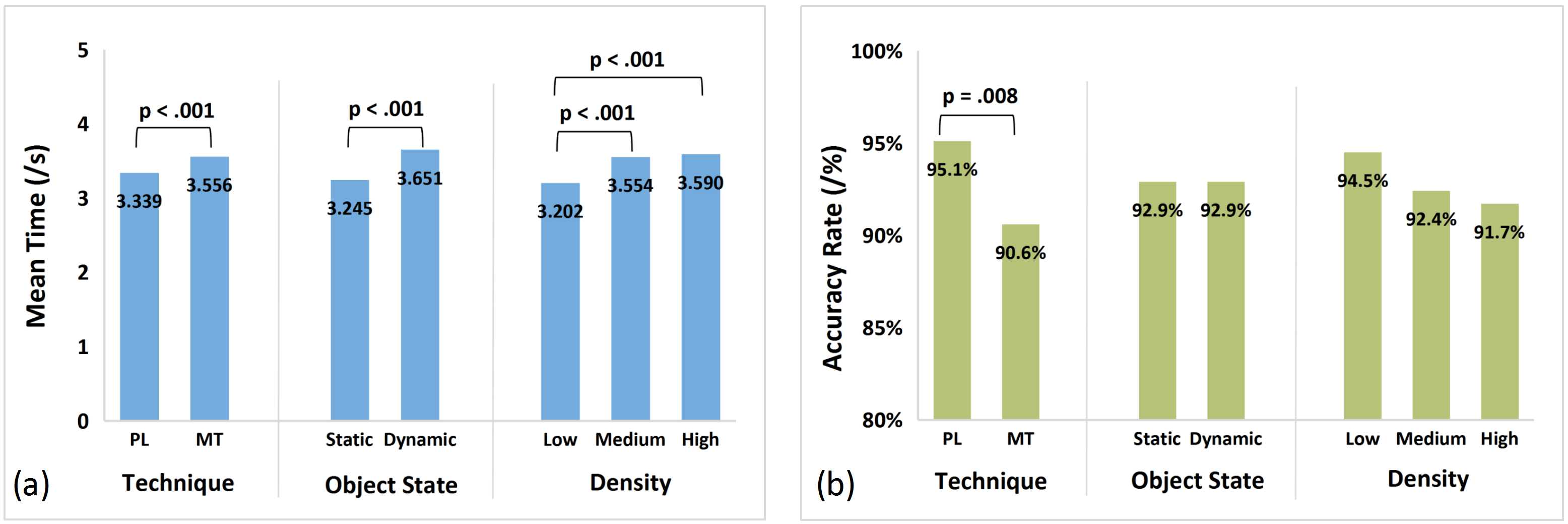}
  \caption{Completion Time (a) and Accuracy Rate (b) according to Technique, Object State, and Density.}
  \label{Fig4}
\end{figure*}

\subsection{Results}
For the objective data analysis (completion time and accuracy rate), we employed three-way repeated ANOVAs with an alpha value of 0.05 to determine any differences across conditions and followed by pair-wise comparisons with Bonferroni correction for the data with a significant difference. We used a Greenhouse-Geisser adjustment to correct for violations of the sphericity assumption. We reported effect sizes ($\eta$$_p$$^2$). For data from the subjective questionnaires that were ordinal, non-parametric data  (e.g., subjective ratings or rankings), we applied Wilcoxon signed-rank test to look for differences. For simplicity, M and SD are used to denote mean and standard deviation values.
\subsubsection{Task Performance}
Participants' task completion time and accuracy rate were collected to assess their performance. We recorded the time taken by paired participants to complete each trial. Accuracy rate was measured by the number of correct trials among the total number of trials. Figure~\ref{Fig3} shows mean time and accuracy rate for each condition (2 Technique × 2 Object State × 3 Density). 

\paragraph{\textbf{\textit{Completion Time}}} 
We found an interaction effect between Technique × Object State (F$_{1, 95}$ = 3.993, p = .049, $\eta$$_p$$^2$ = .040), Technique × Density (F$_{2, 190}$ = 7.461, p = .001, $\eta$$_p$$^2$ = .073), Technique × Object State × Density (F$_{2, 190}$ = 4.065, p = .019, $\eta$$_p$$^2$ = .041) on completion time, but no significance was found between Object State × Density (p = .758). For Technique × Object State, pairwise comparisons revealed that PL\_Dynamic took significantly less time than MT\_Dynamic (p = .001). PL\_Static also took less time than MT\_Static, but no significant difference was found (p = .149). In addition, both PL and MT techniques took significantly less time in static condition than that in dynamic one (both p < .001).

For Technique × Density, PL\_Medium was significantly faster than MT\_Medium (p < .001). Although we noticed that PL\_High and PL\_Low were respectively faster than MT\_High and MT\_Low, no significant difference was found (both p > .05). Besides, PL\_High was significantly slower than PL\_Medium (p = .027) and PL\_Low (p < .001). MT\_Low was significantly faster than MT\_High (p = .001) and MT\_Medium (p < .001). For Technique × Object State × Density, PL\_Dynamic\_Medium took less time than MT\_Dynamic\_Medium (p < .001). For other conditions, although using PL was faster than MT, there was not any significant difference (p > .05).

\begin{table*}[b]
\centering
  \caption{Means (Standard Deviations) of accuracy rate on different conditions.}
  \label{tab2}
  \begin{tabular}{m{1cm}<{\centering}  p{1cm}   p{1cm}  p{1cm} p{2cm}  p{1cm}   p{1cm}  p{1cm} p{2cm}}
    \toprule
    \multirow{2}*{}& \multicolumn{4}{c}{\textbf{Static}} & \multicolumn{4}{c}{\textbf{Dynamic}}\\
    \cline{2-9}
    &Low&Medium&High& {\textbf{Overall Mean} } &Low&Medium&High& {\textbf{Overall Mean}}\\
     \hline
     \textbf{PL}& \multicolumn{1}{m{1cm}}{97.9\% (14.4\%)}&\multicolumn{1}{m{1cm}}{93.8\% (24.3\%)}& \multicolumn{1}{m{1cm}}{92.7\% (26.1\%)}& \multicolumn{1}{m{2cm}}{94.8\% (22.3\%)}& \multicolumn{1}{m{1cm}}{95.8\% (20.1\%)}&\multicolumn{1}{m{1cm}}{95.8\% (20.1\%)}&\multicolumn{1}{m{1cm}}{94.8\% (22.3\%)}&\multicolumn{1}{m{2cm}}{95.5\% (20.8\%)}\\
      \hline
       \textbf{MT}& \multicolumn{1}{m{1cm}}{93.8\% (24.3\%)}&\multicolumn{1}{m{1cm}}{87.5\% (33.2\%)}& \multicolumn{1}{m{1cm}}{91.7\% (27.8\%)}& \multicolumn{1}{m{2cm}}{91.0\% (28.7\%)}& \multicolumn{1}{m{1cm}}{90.6\% (29.3\%)}&\multicolumn{1}{m{1cm}}{92.7\% (26.1\%)}&\multicolumn{1}{m{1cm}}{87.5\% (33.2\%)}&\multicolumn{1}{m{2cm}}{90.3\% (29.7\%)}\\
  \bottomrule
\end{tabular}
\end{table*}
\begin{figure*}[b]
  \centering
  \includegraphics[width=1\textwidth]{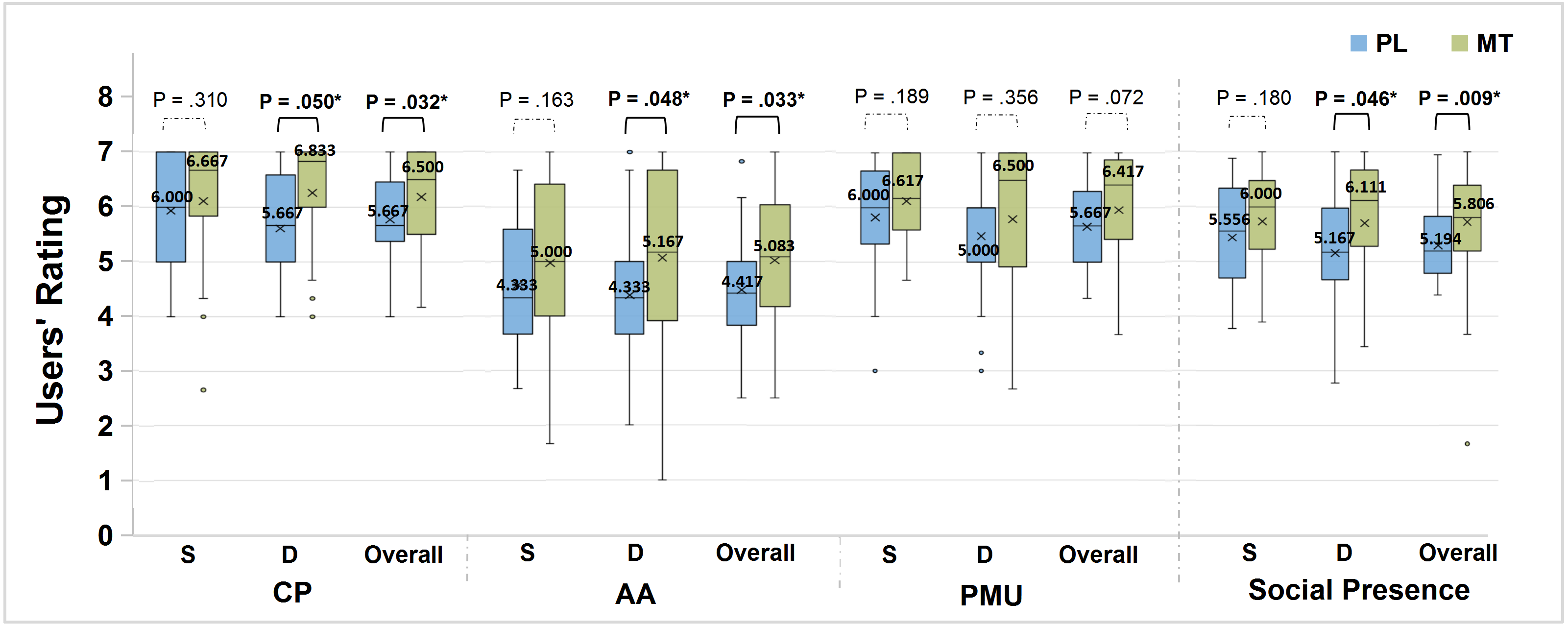}
  \caption{Users' ratings on Social presence in all conditions (2 Techniques × 2 Object States); Significance results are highlighted in \textbf{Bold}; \textbf{PL}: Pointing Line, \textbf{MT}: Moving Track; \textbf{S}: Static, \textbf{D}: Dynamic; Subscales, \textbf{CP}: Co-presence, \textbf{AA}: Attention Allocation, and \textbf{PMU}: Perceived Message Understanding.}
  \label{Fig5}
\end{figure*}

In addition, there was a significant effect of Technique (F$_{1, 95}$ = 14.559, p < .001, $\eta$$_p$$^2$ = .133), Object State (F$_{1, 95}$ = 33.360, p < .001, $\eta$$_p$$^2$ = .260), and Density (F$_{2, 190}$ = 15.303, p < .001, $\eta$$_p$$^2$ = .139) on completion time. As shown in Figure~\ref{Fig4}a, post-hoc analyses indicated that using PL was significantly faster than using MT. Besides, time spent on trials with Static objects was significantly lower than with Dynamic ones. Post-hoc analyses showed that completing High and Medium density trials took significantly longer time than Low density trials (both p < .001). We also found an increasing trend of the mean time spent on Low, Medium and High density tasks. More details of the main results on completion time are provided in the table in Appendix \ref{APP1}.

\paragraph{\textbf{\textit{Accuracy Rate}}} Overall, we found a high average accuracy rate for all conditions (M = 92.8\%, SD = .257) in all trials (see Figure~\ref{Fig3} (Right)). A further analysis showed that there was no significant interaction effect of Technique × Object State (p = .688), Technique × Density (p = .988), and Technique × Object State × Density (p = .410) on accuracy rate. When looking at the descriptive data, we found PL\_Dynamic and PL\_Static got higher rates than MT\_Dynamic and MT\_Static, respectively. Besides, for mean results of accuracy rate, PL performed better than MT in static trials, while MT performed better in dynamic trials. For Density, the accuracy rate of two techniques gradually decreased with the increase in density. In addition, from Figure~\ref{Fig3} (Right), we can see that PL led to a higher rate than MT in general in any density condition. More details of all conditions on accuracy rate are shown in Table \ref{tab2}.

Besides, we found that there was a significant effect of Technique on accuracy rate (F$_{1, 95}$ = 7.355, p = .008, $\eta$$_p$$^2$ = .072). As shown in Figure~\ref{Fig4}b, Post-hoc analyses showed that participants with PL got significantly higher accuracy rate than ones with MT. In addition, we noticed that the rate decreased gradually with the increase in density of objects. The Static and Dynamic trials led to the same rate. However, we did not find any significant difference of Object State and Density (all p > .05).
\subsubsection{User Experience}
Participants' collaboration and user experience were quantified using the data from post-experiment questionnaires that contained Likert-scale questions. We collected three sets of subjective data: Social Presence, System Usability Scale, and User Preference.

\paragraph{\textbf{\textit{Social Presence}}} We adapted the Social Presence Questionnaire \cite{harms2004internal} according to  the nature of our trials. 3 sub-scales, Co-presence (CP), Attention Allocation (AA) and Perceived Message Understanding (PMU), were used and consisted of 9 rating items on a 7-point Likert scale (1: Strongly Disagree $\sim$ 7: Strongly Agree). Overall, participants had a high feeling of social presence when interacting in AR (M = 5.507, SD = .895). Besides, we found that MT was rated higher than PL in all conditions for median scores (see Figure~\ref{Fig5}). A Wilcoxon signed-rank test revealed that there was a significant interaction effect of social presence between PL and MT in dynamic trials (Z = -1.993, p = .046) but no interaction significance was found in static trials  (p = .180). Besides, there was a significant effect of Technique on social presence (Z = -2.618, p = .009). Pairwise comparisons revealed that overall MT got a significantly higher score than PL. 

For the subscales, the Wilcoxon signed-rank test revealed a significant difference of CP on PL and MT (Z = -2.142, p = .032). Post-hoc tests showed that MT got a significantly higher score than PL (Z = -1.961, p = .050) in Dynamic trials but there was no significance in Static trials (p = .310). For AA, we found a significant difference between PL and MT (Z = -2.128, p = .033). MT got a significantly higher score than MT in Dynamic trials (Z = -1.976, p = .048), but there was no significance in Static trials (p = .163). We did not find any significance on the PMU subcale (all p > .05).

\begin{figure}[h]
\centering 
\begin{minipage}[b]{0.45\textwidth} 
\centering
\includegraphics[width=1\textwidth]{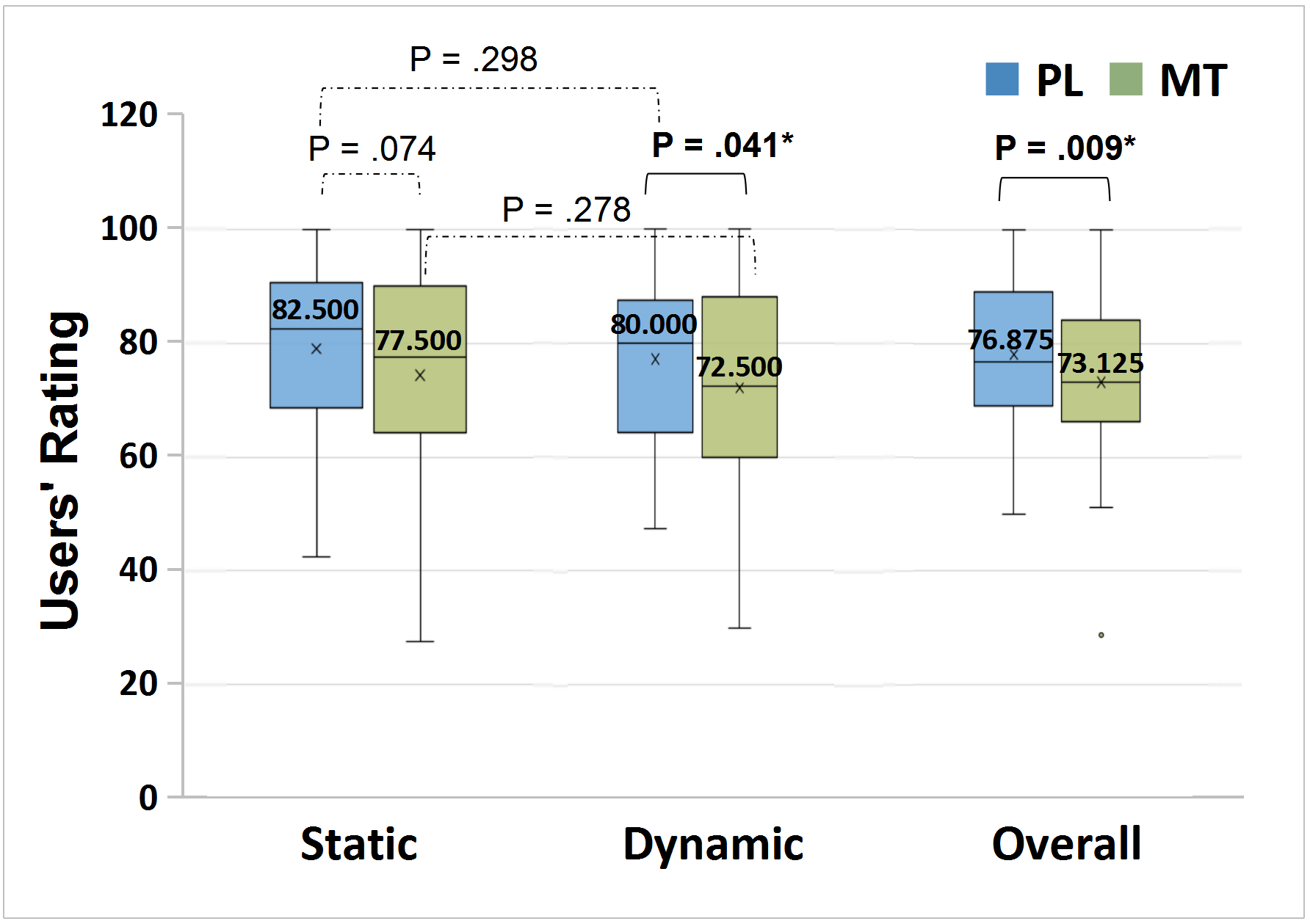} 
\caption{Users' ratings of SUS on the technique's usability with significance (*: statistically significant); SUS score $\in$ [0, 100], the higher, the better.}
\label{Fig6}
\end{minipage}
\begin{minipage}[b]{0.5\textwidth} 
\centering 
\includegraphics[width=0.94\textwidth]{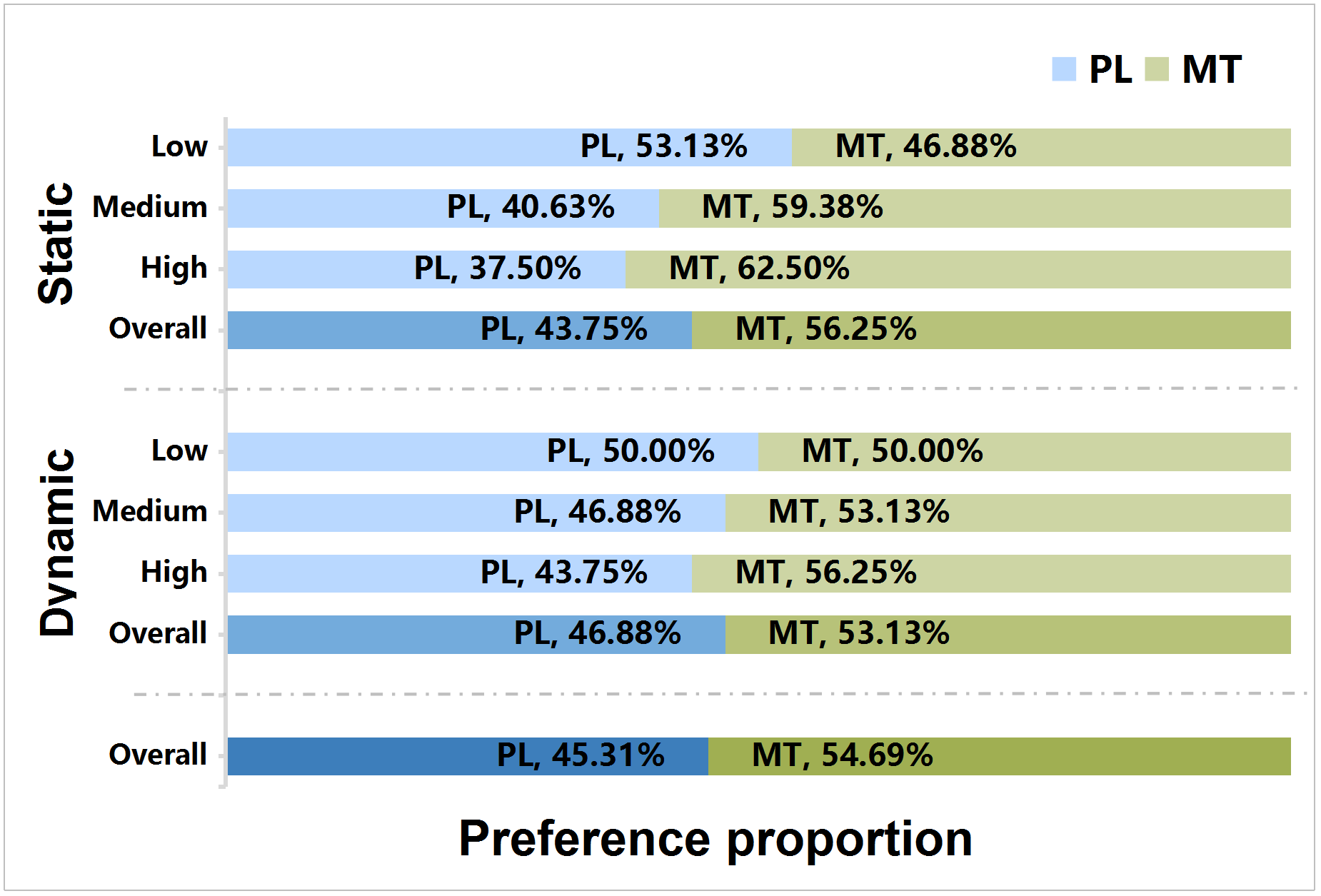}
\caption{User preference of the two visual techniques.}
\label{Fig7}
\end{minipage}
\end{figure}

\paragraph{\textbf{\textit{System Usability Scale (SUS)}}} SUS \cite{brooke1996sus} was selected to measure and evaluate the participants’ responses towards the usability of the two visual techniques. It consisted of 10 items using  5-point Likert scales (1: Strongly Disagree $\sim$ 5: Strongly Agree). Participants gave an average of 78.162 (SD = 12.958) and 73.235 (SD = 14.414) for PL and MT, respectively. Figure~\ref{Fig6} shows the users' rating with median and significance results for all conditions. A Wilcoxon signed-rank test showed that there was a significant difference between PL and MT (Z = -2.611, p = .009). Pairwise comparisons showed a significant effect between PL and MT in Dynamic trials (Z = -2.042, p = .041). No other significance was found (all p > .05).

\paragraph{\textbf{\textit{User Preference}}} At the end of the experiment, we asked participants to choose their preferred technique for different types of trials (see Figure~\ref{Fig7}). Overall, MT (54.69\%) was more favored by participants than PL (45.31\%). For the Static\_Low and Dynamic\_Low trials, the proportion of MT was not higher than PL. Except for these 2 types of trials, MT got always a higher number of votes. One interesting finding was that both in Static and Dynamic trials, the higher the density of objects, more participants would choose MT.

\section{DISCUSSION}
In this paper, we explored the use of two visual techniques (Pointing Line (PL) and Moving Track (MT)) for providing awareness cues during collaborative exploration in AR systems for static and dynamic objects and with different levels of object density (Low, Medium and High). Overall, results from our user study indicated that these visual cues provided benefits. We next discuss the findings in more detail.
\subsection{User Performance and Usability: PL > MT}
When completing the trials, the density and state of objects affected the participants' performance. This was expected and our results verified that the time needed for selecting dynamic objects was significantly longer while the accuracy was lower, which supports \textit{\textbf{H1}}. The accuracy rate decreased with the increase of object density, which also led to increase the time spent on the tasks. In particular, the time spent on lower density trials was significantly less than on higher density trials, which is in line with our expectations and also aligns with \textit{\textbf{H2}}.

Our results show that PL led to significantly higher performance than MT, which aligns with \textit{\textbf{H3}}. This result aligns with the findings from Kim et al. \cite{kim2019evaluating} who found that a pointer could be a main visual communication cue with fast completion time. On the other hand, other studies \cite{fussell2004gestures,kim2013study} found that users completed assembly tasks faster with drawing sketch (moving track) cues than with pointing cues, which differs from our results. However, this difference can be explained with reference to the nature of the tasks in the 2 experiments. For annotations, once drawn, the moving track cues remain in the shared task space so the information is available until it is erased. This can be beneficial for assembling tasks as in the cited experiment. In our scenario, instead, keeping the moving track cues would add further visual details to an already busy AR environment and, as such, may not be so convenient (e.g., in high density cases). For completion time, specifically, PL contributed to a significant higher efficiency than MT in dynamic trials with medium density. In other conditions, PL still performed better than MT in general. In other words, despite the possibility of PL cues occluding more users' view when objects were motion, this did not lower their performance. One participant said that "\textit{PL was easy to control even if the objects were moving"}, which could have provided some degree of counterbalance. As such, this result shows that regardless of whether objects were moving or not and the density level, PL cues seem more beneficial on task performance.

While the accuracy rate was high for both visual techniques, PL was significantly better than MT, which supports \textit{\textbf{H3}}. For mean values, PL led to higher accuracy than MT in all conditions. It seems that PL was more intuitive when showing the position of the target that one user was watching. Some participants (N = 3) mentioned that since a pointing ray was emitted from the user’s head, they would have control over the start point and orientation of the ray, much like a physical laser pointer. Participants found MT slightly more difficult to use, which resulted in taking significantly longer time to complete the trials. In general, our results show that PL was significantly better than MT on task performance, especially for improving task efficiency in dynamic trials with medium density level of objects. 

Results from SUS show that participants rated PL significantly higher than MT. It was preferred significantly more for dynamic trials. In general, participants found that showing a visual line from the head to a target allowed them to see the direction of pointing and this extra implicit information led to improved performance and higher accuracy. Participants used the words '\textit{intuitive}' and '\textit{natural}' to describe PL but not for MT.  

\subsection{Social Presence and User Preference: PL < MT}
We found that MT yielded significantly higher ratings on social presence than PL overall and, specifically, in dynamic trials, which supports \textit{\textbf{H4}}. In particular, MT was rated significantly higher than PL on Co-presence and Attention Allocation for both overall and in dynamic trials. The visual tail of MT moving behind the cursor seems to have allowed users to focus and encouraged them to predict the end target. By focusing on the path, participants said that they `felt more connected' with the thinking of the other user and could sense their presence better. MT's emphasis on the trajectory path in real-time seems to have enhanced their feeling of social presence with the other user. One user mentioned that "\textit{it is very interesting to follow the movement path}", and another one said that "\textit{I can not only know what the target location is, but also where the focus view starts. This way, I can strongly feel my partner being with me together.}" 

Interestingly, when asked which technique they prefer to use, MT was rated higher than PL in most conditions, except for low density. This will support our \textit{\textbf{H4}}. Participants in general said that they prefer MT over PL but there is a general agreement that PL has a better usability, is more intuitive to use, and can help them perform faster. This also explains the findings from the SUS results. This result shows that when working with another person, the feeling of being together, of being connected (even by using a simple visual cue) with the collaborator is an important factor that affects user experience. Some prior studies seem to support our results. For example, Teo et al. \cite{teo2018hand} reported that all of their participants preferred having visual annotation cues, and also stated that participants felt that the task was easier and more enjoyable with dynamic visual cues. Their findings are aligned with our results. Unlike performance results, these results are more subjective and reflect the emotional side of working with other users, which is an important aspect for collaborative systems.

In short, the results show that PL was better on task performance and usability but MT was rated on social presence and user preference.

\subsection{Design Implications of our Findings}
Our results and findings point to the following implications for the design and use of visual cues in co-located AR that involves pointing tasks:
\begin{itemize}
\item The objective results showed that users performed better with Pointing Lines (PL). Therefore, if the goal is to maximize task performance, especially considering efficiency and accuracy, a technique like the PL could be helpful. 
\item PL is considered easy to control and seems more usable, which can transfer to improved task performance. As such, when ease of control and high usability is important, a PL-based technique could be beneficial, especially when dealing with dynamic objects within a medium level of density (slightly occluded and crowded conditions).
\item Based on users' preference and feedback, it seems that being able to follow the movement of the partner's viewing direction provides a higher focus and attention. In this context, they rated Moving Tracking (MT) higher than PL. Therefore, if the goal is to provide higher collaborative experience and social presence, a technique similar to MT is more suitable.
\end{itemize}

\section{LIMITATIONS AND FUTURE WORK}
This research has some limitations, which can serve as directions for future work. As stated earlier, given the results of our pilot study and our literature review, we did not include a baseline technique as by including one it would unnecessarily lengthen our experiment unnecessarily but without providing additional insights. Future work could involve more variations and different implementations of the two visual techniques, a baseline technique can be helpful to allow pinpointing the effect of more specific aspects of visual techniques for pointing tasks in collaborative scenarios. 

Although the size of our sample population is in line with some publications reporting similar experiments (e.g., \cite{huang2019sharing,piumsomboon2019effects}), the sample size was not very large. However, our data still had enough power and was able to show significant differences across the various conditions of the experiment, leading to some interesting findings. In the future, it will be useful to run a study with a larger sample (that has a more diverse group of participants, including pairs who are familiar and unfamiliar with each other) and see if the same results hold or new insights are found. Users in the current study have specific roles, i.e., one initiates the task and the other follows. It would be interesting to study more complex tasks where users' roles are not clearly specified (and can switch back and forth between roles). Future work may also investigate how visual cues can facilitate communication without a specific task and in more fluid and flexible scenarios).

We did not consider the effect of position arrangement of each pair of users and cases where they are located in separate physical spaces. There may not be a significant effect with MT. However, it is not clear what the effect is with PL because it provides implicit direction information about where the user is looking at. In the future, it will be helpful to explore cases where users are moving and positioned at different locations and orientations and when they are separated physically. 

In this experiment, we did not consider other types of non-visual sensorial cues, like audio and haptic feedback. While the implementation of such cues requires careful design, it is interesting and useful because AR systems are multimodal and suitable for combining several modes of sensorial feedback to create a more immersive work environment. 

\section{CONCLUSION}
In this paper, we evaluated the effect of visual cues on pointing task performance, usability, and social presence in co-located AR collaboration. A user study was conducted by comparing two different visual cues (Pointing Line (PL) and Moving Track (MT)) during collaboration with Static and Dynamic tasks in Low, Medium, and High levels of object density. Based on the results of an experiment following a 2 × 2 × 3 (2 Technique × 2 Object State × 3 Density) within-subjects design with 16 pairs of participants, we found that users with PL cues performed better than with MT cues on task performance. Users were more positive about the usability of PL than MT, especially in dynamic tasks with a medium level of object density. Besides, we found that MT cues were useful for enhancing the sense of social presence and user experience when completing pointing tasks in co-located AR. Overall, our results show that PL was better on task performance and usability, while MT was better on social presence and user preference. With these findings, we discussed the implications for the design and use of visual cues for co-located AR collaboration. 

\begin{acks}
The authors wish to thank the participants for their time and the reviewers for their insightful comments that have helped improve our paper. This research was funded in part by Xi'an Jiaotong-Liverpool University's Key Special Fund (\#KSF-A-03 and \#KSF-A-19) and Research Development Fund (\#RDF-16-02-43), the Natural Science Foundation of Guangdong Province (\#2021A1515012629), and Guangzhou Basic and Applied Basic Foundation (\#202102021131).
\end{acks}

\bibliographystyle{ACM-Reference-Format}
\bibliography{sample-base}

\appendix
\section{Appendix A: Summary Table of Main results of completion time showing significant differences between conditions.}
\label{APP1}

\begin{table*}[h]
  \label{tab1}
  \begin{tabular}{ p{2.9cm}<{\centering}  p{1.8cm}   p{2cm}  p{2cm}   p{4.5cm}}
    \toprule
    \textbf{Variable}&\textbf{Condition}&\textbf{Mean (SD) / s}&\textbf{ANOVA - P}&\textbf{Post-hoc Test Results}\\
    \midrule
    \multirow{2}*{Technique} & PL & 3.339 (1.103)&   \multirow{2}*{\textbf{p < .001***}} &   \multirow{2}*{\textbf{PL  <  MT  (p < .001***)}}\\
   & MT & 3.556 (1.205) \\
  \hline
     \multirow{2}*{Object State} & S & 3.245 (0.960)&  \multirow{2}*{\textbf{p < .001***}} &  \multirow{2}*{\textbf{{S  <  D  (p < .001***)}}}\\
   & D & 3.651 (1.299)\\
  \hline
     \multirow{3}*{Density} & L & 3.202 (1.103) &  \multirow{3}*{\textbf{p < .001***}} & {\textbf{L   <    H    (p < .001***)}}\\
   & M & 3.554 (1.181) & & {\textbf{L  <    M    (p < .001***)}}\\
   & H & 3.587 (1.157) & & M   <   H    (P = .998)\\
  \hline
       \multirow{4}*{\makecell[tc]{Technique\\×\\Object State} }& PL\_S & 3.195 (0.981) &  \multirow{4}*{\textbf{{p = .049*}}}&  \multirow{4}*{ \makecell[tl]{\textbf{PL\_D  <  MT\_D   (p = .001**) }\\ PL\_S   <  MT\_S    (P = .149)}}\\
   & PL\_D & 3.483 (1.197) & & \\
   & MT\_S & 3.294 (0.937) & & \\
   & MT\_D & 3.818 (1.376) & & \\
  \hline
      \multirow{6}*{ \makecell[tc]{Technique\\×\\Density} } & PL\_L & 3.161 (1.114) &  \multirow{6}*{\textbf{{p = .001**}}}&  \multirow{6}*{ \makecell[tl]{{\textbf{PL\_M  <  MT\_M  (p < .001***) }}\\ PL\_H   <  MT\_H   (P = .381) \\ PL\_L    <  MT\_L   (p = .428 )}}\\
   & PL\_M & 3.392 (0.937) & & \\
   & PL\_H & 3.551 (1.200) & & \\
   & MT\_L & 3.243 (1.094) & & \\
   & MT\_M & 3.717 (1.366) & & \\
   & MT\_H & 3.622 (1.115) & & \\
  \hline
      \multirow{12}*{ \makecell[tc]{Technique\\×\\Object State\\×\\Density} } & PL\_S\_L & 2.951 (0.894) &  \multirow{12}*{\textbf{{p = .019*}}}&  \multirow{12}*{ \makecell[tl]{\textbf{{PL\_D\_M  <  MT\_D\_M   (p < .001***)}}\\
      PL\_D\_H  <   MT\_D\_H   (p = .494)\\
      PL\_D\_L   <   MT\_D\_L   (p = .579)\\
      PL\_S\_H   <   MT\_S\_H   (p = .588)\\
      PL\_S\_M  <   MT\_S\_M   (p = .103)\\
      PL\_S\_L   <   MT\_S\_L    (p = .516)}}\\
   & PL\_S\_M & 3.250 (1.006) & & \\
   & PL\_S\_H & 3.385 (0.998) & & \\
   & PL\_D\_L & 3.372 (1.267) & & \\
   & PL\_D\_M & 3.360 (0.894) & & \\
   & PL\_D\_H & 3.718 (1.357) & & \\
   & MT\_S\_L & 3.024 (0.909) & & \\
   & MT\_S\_M & 3.424 (0.982) & & \\
   & MT\_S\_H & 3.435 (0.866) & & \\
   & MT\_D\_L & 3.461 (1.218) & & \\
   & MT\_D\_M & 4.184 (1.516) & & \\
   & MT\_D\_H & 3.809 (1.295) & & \\

  \bottomrule
 {\footnotesize {\makecell[tl]{ \textbf{Note}: Level of significance: \textbf{*}<0.05, \textbf{**} <0.01, and \textbf{***} <0.001; Significant results are highlighted in \textbf{Bold}; \textbf{PL}: Pointing Line, \textbf{MT}: Moving Track; \\ \textbf{S}: Static, \textbf{D}: Dynamic; \textbf{L}: Low, \textbf{M}: Medium, \textbf{H}: High.}}}
\end{tabular}
\end{table*}

\end{document}